# Empirical relationship as a stepping-stone to theory

Marijan Ribarič and Luka Šušteršič
Jožef Stefan Institute, Ljubljana, Slovenia
E-mail: marjan.ribaric@masicom.net

## Abstract

Empirical equations are mathematical equations whose free parameters we have to specify by a given set of experimental data about a particular state to obtain formulas for predicting other data about this state. By Ockham's razor problem-solving principle, it is always understood that such empirical equations should be constructed parsimoniously, thereby trying to achieve the required accuracy with fewest free parameters. In constructing empirical equations, we use heuristic techniques of computational physics.

We consider parsimonious construction of empirical equations, to promote interest in them as a stepping-stone model to the physical law. To this end, we provide a variety of historical examples and simulate a parsimonious empirical calculation of Planck's law, and of van der Waals' equation. Thereby we provide **a**) Empirical forms of Planck's law, and **b**) Collation of verified symmetries and catastrophes-like properties of empirical P-ρ-T surface of real gases. An empirical equation of state for a real gas should take account of these properties.

**Keywords**: Computational physics, physical laws, catastrophe theory, empirical P(ρ,T) equations, empirical propagators

# Content

## 1. Introduction

Empirical relationships, such as empirical equations, we calculate for predicting results of new experiments from the old ones. Historically, science has used many empirical relationships as preliminary mathematical models, stepping-stones to theories providing physical laws that generalize and extend them, see subsequent Section and [6, 9, 34, 36]. Accordingly, we put forward in [28] the following **Proposition about the empirical equations:**

*An empirical* equation *of outstanding fitting quality may be a key stepping-stone towards a pertinent theory*.

- Parsimonious data fitting is crucial for creating such empirical equations.

### 1.1. Examples

[1st] **Inspirations from the history of calculus.** Both major branches of calculus, differential and integral calculus, provide a wide, invention inspiring supply of empirical relations and ideas that have been eventually generalized and extended to packages of techniques, which were then subsumed by lemmas and theorems developed theoretically in a rigorous an systematic way, see [41]. They suggest how one can calculate (guess) novel something, using a special way and all facts that one can find.

[2nd] **Pythagorean Theorem** is actually a Babylonian empirical relation about four thousand years old. It states that the area of the square whose side is the hypotenuse of the right angle (with length $c$) equals the sum of the areas of the squares on the other two sides of the triangle (with lengths $a$ and $b$). Thus, there is the Pythagorean equation of this theorem

$$a^2 + b^2 = c^2, \qquad (1)$$

which was actually in use for over a thousand years before Pythagoras.

[3rd] **Ptolemaic system** espoused in the 2nd century AD, has evolved over the Copernican system to Newton's law of universal gravitation, see [1].

[4th] **Hooke's law** stated in 1676 has been generalized to a tensor expression to study deformations of various materials, see [2].

[5th] **Ohm's law** of 1827 is still an extremely useful empirical equation in electrical engineering, see [3].

[6th] **Stefan-Boltzmann-Law** deduced in 1879 by Josef Stefan empirically, his student Ludwig Edward Boltzmann theoretically derived in 1884, see [42].

[7th] **Balmer's formula** of 1885, predicting the spectral lines of the hydrogen atom, and its generalization, Rydberg's formula of 1888 that predicts the spectral lines of hydrogen-like atoms were incorporated into the Bohr's model in 1913, a stepping-stone to quantum mechanics, see [4, 5].

## 1.2. Overview

How empirical observations of physical behavior and empirical laws determine the form and content of physics and its theoretical structures, Cook [6] has considered in straightforward, nontechnical style. Dealing with basic aspects of the general framework of physics, he provides a lucid examination of issues of fundamental importance in the empirical and metaphysical foundations of physics.

**In Chapter 2,** as a supplement to Cook's work we simulate calculations of Planck's law through empirical equations.We collate the issues and activities and briefly describe the area and scope of work that are necessary for this. There are two crucial facts: there is always only a finite number of experimental data, and they are not accurate; to avoid unilluminating complications, we will consider only the averaged experimental data.

**In Chapter 3**, we consider the empirical equation of state for a real gas.

**In Chapter 4**, we introduce empirical propagators. Chapter 2 gives the basic examples, whereas Ch. 3, and Ch. 4 provide independently additional ones..

**In Chapter 5**, there are concluding remarks with supplementary information; we mention some basic characteristics of the underlying natural processes that are relevant to making empirical equations.

- We will denote by the adjective P(n) an empirical equation that has n free parameters, and by an additional *global or general* when we use it for predicting all possible experimental results of a particular state.
- As ansatze for an empirical equation, we will use various mathematical equations whose parameters we interpret as free ones.
- As we do not presume that an empirical equation is strictly accurate, we cannot prove or falsify it by the experimental data. Data can only show how compatible it is with them. To paraphrase an old saying:

"*The proof of the empirical equation is in its usage*".

### 1.3. Parsimonious Approach

Empirical equations are mathematical equations whose free parameters we have to specify by a given set of experimental data about a particular state to obtain formulas for predicting other data about this state; a variable or an index may specify the relevant state. By Ockham's razor problem-solving principle, it is always understood that such empirical equations should be constructed parsimoniously [26], thereby trying to achieve the required accuracy with fewest free parameters. In constructing **e**mpirical equations, we use heuristic techniques of computational physics. There are tree simple illustrative cases of **parsimonious data fitting**:

**i) Selection of Ansatz Parameters.** We are given function $F(x)$ and the n-term power series

$$P_n(x) = c_o + c_1 x + \cdots + c_n x^n \qquad (2)$$

as the P(n+1) empirical ansatz, where it is customary that the initial values of its parameters $c_i = 0$ .To approximate $F(x)$ by $P_n(x)$ we determine the empirical values of its n+1 free parameters $c_i$ by some data-fitting method, e.g. by least square. Thereafter we can create a *parsimonious kind of approximation,* say $E_1(x)$ by ignoring all of the less important parameters of $P_n(x)$, which we consider not to contribute sufficiently to the overall quality of approximation. To this end, we replace their empirical values by the initial ones, thus eliminating them. Thence we are proceeding with

**ii) Iterative improvement.** To improve $E_1(x)$ we take $P_n(x)E_1(x)$ to approximate $F(x)$ and determine the empirical values of n+1 free parameters of $P_n(x)$ to obtain iteratively the second order parsimonious approximation, say $E_2(x)$, by eliminating the less important parameters of $P_n(x)$ by replacing their empirical value by the initial ones. Proceeding this way we determine the highest order parsimonious kind of approximation, say $E_{opt}(x)$ that we cannot improve any more by this kind of iteration.

**iii) Choosing Appropriate Ansatz.** For instance, among P(N+1) rational functions

$$R_{N,n} = (c_o + c_1 x + \cdots + c_n x^n) / \, (1 + c_{n+1} x + \cdots + c_N x^{N-n}) \, , n = 0,1, \ldots \, N, \quad (3)$$

we determine the best of all N+1 ansatze and use it for parsimonious fitting.

## 2. Black-body Radiation

One may consider Planck's law as the most famous and significant empirical equation of theoretical physics, cf [34]. To point out a variety of particular challenges we encounter using experimental data to infer a physical law, we will simulate a calculation of Planck's law through empirical equations, (and thus the Stefan-Boltzmann-Law it implies, see [42]). Planck's law describes the spectral radiance $I(\nu, T)$ of electromagnetic radiation of a black body as a function of frequency $\nu$ and its temperature $T$:

$$I(\nu, T) = 2hc^{-2}\nu^3/(e^{h\nu/kT} - 1)\ . \tag{4}$$

To simulate calculation we consider a set of $J$ distinct black-body states, with $T_j$ denoting the temperature of the state $j = 1,\ldots, J$. Let us observe the spectral radiance of each state at $I$ different frequencies $\nu_i$, $i = 1,\ldots, I$, repeating each measurement $N$ times, cf. [6, Sect.1.2]. We denote the actually measured spectral radiance by $i_j(\nu_i;\ n)$, $n = 1,\ldots, N$, and their averages by

$$i_j(\nu_i) \equiv N^{-1} \textstyle\sum_{n=1}^{N} i_j(\nu_i; n)\ . \tag{5}$$

On having obtained a finite number of averaged experimental data about the spectral radiance of the jth black-body state, say

$$D_j \equiv \{i_j(\nu_i), i = 1, \ldots, I\}\ , \tag{6}$$

we have two problems, which we will now consider in this Chapter:

**[1st]** How to construct an empirical equation by using only a few of the averaged experimental data $i_j(\nu_i)$, and then test it with the rest of them.

**[2nd]** Construction of empirical equations that represent the physical law (Planck's law) that underlies $D_j$, i.e. how to obtain quantitative information about physical law from the given experimental data.

### 2.1. Local Empirical Equations

Experience suggests we start constructing with a proven, linear ansatz

$$E_1(\nu) = c_1 + c_2\nu\ . \tag{7}$$

There are many ways to specify free parameters $c_1$ and $c_2$ by data $i_j(\nu_i)$. Fitting two of them, say $i_j(\nu_1)$ and $i_j(\nu_2)$, we get

$$E_1(\nu) = L(\nu; i_j(\nu), \nu_1, \nu_2), \quad (8)$$

where

$$L(\nu; i_j(\nu), \nu_1, \nu_2) \equiv i_j(\nu_1) + [i_j(\nu_2) - i_j(\nu_1)](\nu - \nu_1)/(\nu_2 - \nu_1)\,. \quad (9)$$

This local, P(2) empirical equation reproduces data $i_j(\nu)$ well if $\nu$ is sufficiently close to $\nu_1$ or to $\nu_2$ [1]. When such linear ansatz cannot reproduce the averaged experimental data well enough, we might try a local, higher-order polynomial ansatz with more free parameters. Nevertheless, if too many free parameters seem to be required, different kind of empirical equations might be more appropriate.

Thus, on plotting $\ln(\nu_i^{-3} i_j(\nu_i))$ versus $\nu_i \to \infty$, we could come to presume that

$$E_\infty(\nu) = \nu^3 \exp[L(\nu; \ln(\nu^{-3} i_j(\nu)), \nu_1, \nu_2)] \quad (10)$$

would be an appropriate, local, P(2) empirical equation for predicting the values of $i_j(\nu)$ for high frequencies $\nu \to \infty$, provided we choose both $\nu_1$ and $\nu_2$ large enough [1].

Similarly, on plotting $\ln(i_j(\nu_i))$ versus $\ln(\nu_i)$, we could presume that the local, P(1) empirical equation

$$E_0(\nu) = i_j(\nu_0)(\nu/\nu_0)^2 \quad (11)$$

will turn out suitable for predicting the values of $i_j(\nu_i)$ as frequencies $\nu \to 0$, provided we chose $\nu_0$ small enough [1].

### 2.2. Lawlike, Empirical Equations

An empirical equation that is entirely specified by n, *empirical constants,* we will denote as a *lawlike, P(n) empirical* equation.

- An approximation to an equation describing a given phenomenon becomes a lawlike P(n) empirical equation when we specify n of its constants empirically.
- As its form suggest, a lawlike, empirical equation might actually approximate an underlying physical law.
- We will combine local empirical equations to obtain a more extensive and parsimonious empirical equations.

---

[1] For simplicity, and to avoid being tedious, we will not formally specify this.

Since the preceding three empirical equations $E_1(\nu,T_j)$, $E_\infty(\nu,T_j)$ and $E_0(\nu,T_j)$ did turn out to be useful for predicting particular averaged experimental data about the radiance of the jth state, we now use them to make local, lawlike, empirical equations, thereby hoping to get parsimoniously the appropriate approximations to the underlying law.

To make them, we make two basic assumptions about the accuracy of the averaged experimental data $D_j$:

[1st] The averaged experimental data $i_j(\nu_i)$ tend toward the values of spectral radiance when the number of repeated measurements $N$ becomes large, i.e.

$$\lim_{N\to\infty} i_j(\nu_i) = I(\nu_i,T_j) \in (0,\infty) \quad \forall\ i,j. \tag{12}$$

[2nd] The maximal relative observational error of data $D_j$ tends to zero as the number $N$ of repeated observations becomes large:

$$\max_{i,j}|i_j(\nu_i) - I(\nu_i,T_j)|/I(\nu_i,T_j) \to 0 \qquad as \quad N\to\infty\,. \tag{13}$$

In addition, we make also the following four assumptions about the properties of the spectral radiance $I(\nu,T)$:

[3rd] There is the limit

$$c_{1j} \equiv \lim_{\nu_2\to\nu_1}[I(\nu_2,T_j) - I(\nu_1,T_j)]/(\nu_2-\nu_1)\ \forall\ \nu_1 > 0. \tag{14}$$

[4th] There is the limit

$$c_{2j} \equiv \lim_{\nu\to 0} I(\nu,T_j)/\nu^2. \tag{15}$$

[5th] There is the limit

$$c_{\infty j} \equiv -\lim_{\nu\to\infty}\nu^{-1}\ln[I(\nu,T_j)/\nu^3]\,. \tag{16}$$

[6th] There is the limit

$$c_{3j} \equiv \lim_{\nu\to\infty}\nu^{-3}e^{c_{\infty j}\nu}i_j(\nu)\,. \tag{17}$$

These limits define four theoretical constants, which describe the intensive physical properties of the jth black-body state. To determine their values, we would need an infinite number of averaged experimental data, but the number of available ones is always finite. When using mathematical formulas with theoretical constants as ansatze to empirical formulas about data $D_j$, we have to replace the theoretical constants with empirical ones, i.e. with estimated values of them, based on the $IN$ available experimental data. Metrological methods provide various estimates for empirical constants. We assume herein that they are getting entirely accurate with larger numbers $IN$ of data [1], cf. (12) and (13).

Inspecting the empirical values of the constant $c_{3j}$ and of the products $c_{2j}c_{\infty j}$ for the given states, we could come to presume that the values of $c_{3j}$ and $c_{2j}c_{\infty j}$ equal the same state-independent value, say $c_3$, i.e.

$$c_{2j}c_{\infty j} = c_{3j} = c_3 \quad \forall \quad j = 1,2,\dots,J\,. \tag{18}$$

### 2.3. Local, Lawlike, Empirical Equations

**a) Linear case.** The P(2) empirical equation $E_1(\nu, T_j)$ suggests the following ansatz for a local, lawlike, P(2) empirical equation

$$L_1(\nu, T_j) = I(\nu_1, T_j) + c_{1j}(\nu - \nu_1) \quad as \quad \nu \to \nu_1\,. \tag{19}$$

Testing it on the available data $D_j$, we get an empirical value of the constant $c_{1j}$, and an estimate of values $|\nu - \nu_1|$ where this local empirical equation is still useful. Hooke's law and Ohm's law are examples of such linear equations.

**b) Low frequencies.** The usefulness of the empirical equation $E_0(\nu, T_j)$ indicates that the asymptotic behavior of the spectral radiance $I(\nu, T_j)$ at low frequencies will be well described by the following P(1) ansatz

$$L_0(\nu, T_j) = c_{2j}\nu^2 \quad as \quad \nu \to 0 \tag{20}$$

if the empirical value of $c_{2j}$ is accurate enough. The RHS(20) equals the Rayleigh-Jeans formula.

**c) High frequencies.** The usefulness of the empirical formula $E_\infty(\nu, T_j)$ for predicting the values of $i_j(\nu_i)$ for high frequencies suggests that the ansatz

$$L_\infty(\nu, T_j) = c_3\nu^3 e^{-c_{\infty j}\nu} \quad as \quad \nu \to \infty \tag{21}$$

would result in an adequate local, lawlike, P(2) empirical formula for predicting the high frequency, asymptotic behavior of $I(\nu, T_j)$. The RHS(21) equals Wien's law and it predicts also the correct limiting value of spectral radiance for $\nu = 0$, but does not imply the correct asymptotic behavior (20) for low frequencies.

## 2.4. Global, Lawlike, Empirical Equations

As the ansatze $L_1$, $L_0$ , and $L_\infty$ for local, lawlike, empirical formulas have turned out satisfactory for predicting the averaged experimental data $D_j$ locally, we will combine them to construct a global, lawlike, empirical formula, intended for predicting all frequencies of the jth state. They suggest the assumption

[7th] Spectral radiance is a non-negative, analytic function in a vicinity of any positive frequency.

After certain juggling, inspired by the Padé approximants we construct the following ansatz for the lawlike, P(3) empirical equation that unifies by an interpolation the ansatze $L_0$ and $L_\infty$, for local, lawlike, empirical equations and agrees with the [7th] assumption:

$$U(\nu; c_3, c_{\infty j}, c_j) = c_3\nu^2(c_j + \nu)/(e^{c_{\infty j}\nu} - 1 + c_j c_{\infty j}) \qquad (22)$$

where $c_j$ is a heuristic, non-negative empirical constant. On testing this ansatz with $c_j = 0$, we find that we can represents data $D_j$ well enough for any state if the total number of experimental data, $IJN$ is large enough [1]. Thus we could presume that $U(\nu; c_3, c_{\infty j}, 0)$ represents Planck's law for the radiance of the jth black-body state; it is specified by its two physical properties $c_3$ and $c_{\infty j}$. Thus, $U(\nu; c_3, c_{\infty j}, 0)$ is an empirical, P(2) form of Planck's law for the radiance of the jth black-body state.

## 2.5. An Empirical Form of Planck's Law

On measuring the temperatures $T_j$ of various states and plotting $T_j^{-1}$ versus empirical values of $c_{\infty j}$, we could come to presume that there is such a positive constant $c_\infty$ that

$$c_{\infty j} = c_\infty T_j^{-1} \quad \forall\ j = 1,2, \dots, J\ . \qquad (23)$$

Thence we could put forward the following lawlike, P(2) equation

$$U(\nu; c_3, c_\infty/T, 0) \qquad (24)$$

as an empirical Planck's law that has two empirical constants $c_3$ and $c_\infty$ .

- In the following three subsections, we comment on some relevant issues when constructing Planck's law empirically.

### 2.6. Identifying the States

The black-body states are identified by the index $j$, giving the order in which they were researched. In principle, we could identify the black-body states by their physical properties. So, were they known to us, we could use the values of $c_{\infty j}$ that are inversely proportional to temperature, or the values of spectral radiance for a specific frequency $\nu_m$. Moreover, we could make do also with estimates of $c_{\infty j}$ or $I(\nu_{\mathrm{m}}, T_j)$, were they sufficiently accurate not to overlap for the given states.

### 2.7. Qualitative Properties of Planck's Law

Planck's law (4) is such that

$$I(-\nu, -T) = -I(\nu, T)\,, \qquad I(T\nu, T) = T^3 I(\nu, 1)\,, \tag{25}$$

and $I(\nu, T)$ is an analytic function of the complex variable $\nu$ everywhere, but at $h\nu = 2n\pi kTi$, $n = 1$, 2,... where it has a first-order pole, and at $|\nu| = \infty$ where it has an essential singularity. However, we cannot infer these properties directly from experimental data.

Note that only the constant function is analytic in the whole complex plane. Therefore, an analytic function is, up to a constant, uniquely determined by its singularities. An estimate of a singular point of $I(\nu, T)$ close to a point on the positive real axis we can obtain from the slow convergence of the Taylor expansion at this point.

Were a physical law an even function of $\nu$, this property would not be directly evident from experimental data, though the parsimoniously minded equations that presume this property would be more efficient, particularly near $\nu = 0$. The asymptote $L_0(\nu, T_j)$ is compatible with the hypothesis that Planck's law is an even function of frequency, but not the asymptote $L_\infty(\nu, T_j)$.

### 2.8. The Time for Pondering Is Limited

The maximal relative observational error of the available data $D_j$ is in general always present, though we expect it to get smaller as we increase the number $N$ of repetitions. To obtain $J$ data sets $D_j$, each containing $I$ data, we must perform $NIJ$ measurements. And each one takes some amount of time, say at least $t_m$. However, there is obviously an upper limit, say $T_M$, with the amount of time for observing any given physical phenomena. Thus, the total number of possible measurements is smaller than $T_M/t_m$ . Thus, for the experiments we must have the available time

$$T_M > NIJt_m \,, \tag{26}$$

which limits the number $NIJ$ of measurements. Thus,

**A)** When $T_m < t_m$, the phenomenon is practically unobservable. An example is provided by the black-body radiation of visual frequencies at room temperatures. So parts of Planck's law are practically untestable, see [7].

**B)** For any physical law $L(x)$, there is an infinite number of alternatives $L_a(x) = L(x)(1 + \varepsilon\phi(x))$; and if $\phi(x)$ is bounded, the relative difference

$$|(L(x) - L_a(x))/L(x)| \leq |\varepsilon| \sup|\phi(x)| \tag{27}$$

is arbitrarily small for sufficiently small $|\varepsilon|$. Therefore, when relative differences between predictions of two physical laws are sufficiently small, we will never be able to tell them apart experimentally. Such a problem occurs with the empirical form of Planck's law (24), since the relative difference between $U(\nu; c_3, c_\infty/T, c_j)$ and $U(\nu; c_3, c_\infty/T, 0)$ tends to zero uniformly as the constant $c_j \to 0$. Thus, we will never be able to falsify experimentally the assumption that $c_j$ is actually a very small positive physical constant. It just seems expedient to choose $c_j = 0$.

Such is the case with the mass of the photon, where for some theoretical calculations it is convenient to limit the photon mass to zero only in their final stage, though many theoretical considerations take as their basic presumption that photons have no mass [11]. However, $U(\nu; c_3, c_\infty/T, c_j \neq 0)$ might turn out to be a theoretically significant modification of Planck's law. However, the Stefan-Boltzmann-Law implies that $c_j = 0$. This shows how circumstantial evidence is also important when calculating empirical relations.

## 3. Empirical Equation of State for a Real Gas

### A. van der Walls' Equation

We feel that to get ideas how to make an empirical equation of state for a real gas, it make sense to consider properties and construction of present equations of state. The van der Walls' equation is an essential, conceptual stepping-stone to equations of state for gases and liquids. It has various derivations; see [8] and Thijssen [27], which considers construction of the equation of state in through molecular dynamics simulations. We now recall van der Walls' procedure of 1873:

The ideal gas laws by Boyle-Marriott and Gay-Lussac,

$$P = k_1 T \quad \text{and} \quad P\,V = k_2\,, \tag{28}$$

are relating pressure P, temperature T, and volume V. They suggest that we start constructing empirical equations of state for a real gas by P(1) ansatz

$$P\,V = c_1 T \tag{29}$$

with free parameter $c_1$ . The obtained experimental results would support the hypothetical ideal gas equation of state

$$PV_m = R\,T \tag{30}$$

where $V_m$ is the molar volume and $R$ an empirical gas constant. Following van der Walls, we proceed calculation by the P(3) ansatz with free parameters $c_1$, $c_2$, and $c_3$:

$$(P + c_2 V_m^{-2})(V_m + c_3) = c_1 T. \tag{31}$$

Thus, we derive van der Walls' equation

$$(\mathrm{P} + aV_m^{-2})(V_\mathrm{m} - b) = RT \tag{32}$$

where $a$ and $b$ are empirical gas constants.

The van der Walls' equation has critical point of 2nd order with, critical temperature $T_\mathrm{c} = 8a/27bR$, critical density $\rho_\mathrm{c} = 1/3\mathrm{b}$, and critical preasure $P_\mathrm{c} = a/27b^2$. The following form is of interest:

$$P(x) = P_\mathrm{c}[8(\mathrm{x}+1)\mathrm{T}/T_\mathrm{c} + 3\mathrm{x}^3 - 9\mathrm{x} - 6]/(2-\mathrm{x})\,,\; x = \rho/\rho_\mathrm{c} - 1, \tag{33}$$

it shows that the isothermal pressure density dependences of van der Walls' gas exhibit the Riemann-Hugoniot catastrophe.

Let us sum up some

*Mathematical properties of van der Walls' equation:*

1. $\mathrm{P}(\rho, T)$ is a rational function of density $\rho \geq 0$.
2. The degree of the numerator is greater than that of the denominator.
3. $\mathrm{P}(\rho, T) = 0$ at $\rho = 0$ .
4. The denominator of $\mathrm{P}(\rho, T)$ has zero at $\rho = b^{-1}$.
5. $\mathrm{P}(\rho, T)$ is a is a linear function of temperature T.
6. At critical temperature $T_c$ , there is critical density $\rho_c = 1/3\mathrm{b}$ , and first two derivative of pressure P with respect to density $\rho$ are zero.
7. $\mathrm{P}(\rho, T)$ exhibits the Riemann-Hugoniot catastrophe.

## B. Rational Function and Elementary Catastrophes

Thom [31] theoretically grounded in 1960s and Schulman and Revzen [30] elaborated in detail in 1970s that the theory of catastrophes should be applicable to the study of phase transitions in the non-critical region. Thence van der Walls' equation had inspired various equations of state with a rational function to model an elementary catastrophe with temperature as the controlling parameter [12]. To this end, we use the rational function

$$\mathrm{R(x) = N(x)/D(x)} \qquad (34)$$

as the equation of state, where numerator and denominator:

$$N(x) = a_0 + a_1 x + a_2 x^2 + a_3 x^3 + \cdots a_{d_N} x^{d_N} ,$$
$$D(x) = 1 + b_1 x + b_2 x^2 + b_3 x^3 + \cdots b_{d_D} x^{d_D} . \qquad (35)$$

The Taylor polynomial of degree 6 for rational function R(x) is

$$T_6(x) = c_0 + c_1 x + c_2 x^2 + c_3 x^3 + c_4 x^4 + c_5 x^5 + c_6 x^6 \qquad (36)$$

where the polynomial coefficients $c_i$ are determined by the coefficients $a_i$ and $b_i$ of the rational function as follows:

$$c_0 = a_0$$
$$c_1 = a_1 - c_0 b_1$$
$$c_2 = a_2 - c_1 b_1 - c_0 b_2$$
$$c_3 = a_3 - c_2 b_1 - c_1 b_2 - c_0 b_3$$
$$c_4 = a_4 - c_3 b_1 - c_2 b_2 - c_1 b_3 - c_0 b_4$$
$$c_5 = a_5 - c_4 b_1 - c_3 b_2 - c_2 b_3 - c_1 b_4 - c_0 b_5$$
$$c_6 = a_6 - c_5 b_1 - c_4 b_2 - c_3 b_3 - c_2 b_4 - c_1 b_5 - c_0 b_6 . \qquad (37)$$

Thus,

$c_1 = a_1 - a_0 b_1$

$c_2 = a_2 - a_0 b_2 - a_1 b_1 + a_0 b_1^2$

$c_3 = a_3 - a_2 b_1 - a_2 b_1 - a_1 b_2 + a_1 b_1^2 + 2a_0 b_1 b_2 - a_0 b_1^3$

$c_4 = a_4 - a_0 b_4 - a_1 b_3 - a_2 b_2 - a_3 b_1 + a_0 b_2^2 + 2a_0 b_1 b_3 + 2a_1 b_1 b_2 +$
$a_2 b_1^2 - 3\, a_0 b_1^2 b_2 - a_1 b_1^3 + a_0 b_1^4$ . (38)

- The point $x = 0$ is the critical point of the $A_k$-catastrophe, iff

**i)** the continuously variable coefficients $a_i$ and $b_i$ of the rational function R(x) are such that $R^{(0)}(0) \equiv 0$ , $R^{(k)}(0) \equiv 0$ ,

**ii)** there are the critical values $a_{ci}$ and $b_{ci}$ such that

$$R^{(n)}(0) = 0,\ n = 1,2,\dots,(k-1)\text{ , but } R^{(k+1)}(0) \neq 0\text{ ,} \qquad (39)$$

i.e. $c_0 \equiv 0,\ c_{k-1} \equiv 0$ , $and\ c_i = 0\ , i = 0,1,\dots,k-2$, but $c_k \neq 0$ .

Therefore, if we properly vary the parameters $a_i$ and $b_i$ of the rational function R(x) to affect coefficients $c_i$ of the Taylor polynomial, then $x = 0$ is the critical point and the rational function R(x) with $a_0 = 0$ models:

**a)** $A_2$-*Fold catastrophe* with control parameter $c_1$; and$c_2 \equiv 0$.

**b)** $A_3$-*Cusp catastrophe* with control parameters $c_1, c_2$; and $c_3 \equiv 0$.

**c)** $A_4$-*Swallowtail catastrophe*: control parameters $c_1, c_2, c_3$; $and\ c_4 \equiv 0$.

**d)** $A_5$-*Butterfly catastrophe*: control parameters $c_1, c_2, c_3, c_4$; $and\ c_5 \equiv 0$.

### The unfolding polynomial

$$\mathrm{P(x)} \equiv \mathrm{N(x)} - \mathrm{R(0)\,D(x)}, \qquad (40)$$

is such that $P(0) = 0$, and

$R^{(n)}(0) = 0,\ n = 1,2,\dots,(k-1)$ , but $R^k(0) \neq 0$ iff

$$P^{(n)}(0) = 0,\ n = 1,2,\dots,(k-1)\text{ , but } P^k(0) \neq 0\text{ ,} \qquad (41)$$

i.e*. the point* $x = 0$ *is the critical point of order k, both of the rational function R(x) and of its unfolding polynomial* $P(x)$. However, whenever $a_k \equiv 0,\ b_k \equiv 0$, and $a_{k+1} \neq a_0 b_{k+1}$ , then the unfolding polynomial P(x) models the $A_k$-catastrophe with critical point $x = 0$ and its coefficients $(a_n - a_0\, b_n)$, $n = 1,2,\dots,(k-1)$ , are the control parameters.

The $A_2$-*Fold catastrophe* is also named the Riemann-Hugoniot catastrophe.

- *We will name function F(x) as the* $A_k$ *-rational function with the critical point* $x_c$ *if there is a rational function* $R(x)$ *such that*

$$a_k \equiv 0,\ b_k \equiv 0,\ \ a_{k+1} \neq\ a_0 b_{k+1}, \text{and } \mathrm{R(x)} = \mathrm{F}(\mathrm{x} - x_c)\ . \qquad (42)$$

The van der Walls' isothermal pressure density dependences are the $A_2$-rational functions with the critical density $\rho_c = 1/3b$ .

- If a function F(x) is the $A_k$-rational function and also the $A_j$-rational function, then $|k - j| > 1$ !
- Using an $A_k$-rational function with critical point $x_c$ , we can model anyone $A_j$-catastrophe that has $x_c$ as the critical point and the control parameters $(a_i - a_0\, b_i)$, $n = 1,2,\ldots,(j-1)$ if $j$ equals $1,2,\ldots,k-2$, or $k$ .

Therefore, *empirical Padé approximant that is an $A_k$-rational function suggests that the observed phenomenon might exhibit the $A_k$-catastrophe*!

**The Weierstrass theorems**

We can infer some properties of a rational function (the quotient of two polynomials) from the polynomial properties formalized by Weierstrass theorems. We point out some such properties by considering 2nd order polynomial

$$P_2(x) = ax^2 + bx + c : \qquad (43)$$

**1st)** The *second* order polynomial has *two* zeroes, i.e. the quadratic equation $P_2(x) = 0$ has two, *possibly complex* solutions

$$x_\pm = -b \pm \sqrt{b^2 - 4ac}/2a\ . \qquad (44)$$

**2nd)** They depend *continuously* on the polynomial coefficients *a, b*, and *c*.

**3rd)** They can *factorize* the polynomial, i.e. $P_2(x) = a(x - x_-)(x - x_+)$ .

## C. Computational Physics

Parsimonious construction of empirical equations is a typical heuristic, trial and error endeavor of computational physics. We used high performance PC to facilitate the large number of subsequent trials and perturbation models. Take graphical procedures of equations (10) and (11): There is an efficient, such computational *method of osculating functions* that generalizes graphical procedures and discovers more intricate functional dependencies than graph papers can, see [17, 26]. The method we programmed twenty years ago enabled us to evaluate heuristically a particular 32 free parameter function as a potential ansatz for an empirical P(ρ,T) equation. Therefore we may consider the construction of parsimonious empirical equations as an item of the computational physics, cf. [27]. We now point out three such results:

Using the method of investigating experimentally determined functional dependences by osculating curves, cf. [17], the analysis of isothermal pressure-density dependence of the real gases $N_2$, propane, ethylene, $CO_2$, and Xe suggested in[10, 13-16, 18-21, 24] that outside the critical region

**(i) Isothermal pressure density dependence** is both $A_2$ and $A_4$-rational function of density ρ:

$$P(\rho, T) = [a_0 + a_1(\rho-\rho_0) + a_3(\rho-\rho_0)^3 + a_5(\rho-\rho_0)^5 + a_6(\rho-\rho_0)^6 + \cdots]/ [1 + b_1(\rho-\rho_0) + b_3(\rho-\rho_0)^3 + b_5(\rho-\rho_0)^5 + b_6(\rho-\rho_0)^6 + \cdots], \quad (45)$$

where all free parameters $a_{\mathrm{i}}$ and $b_{\mathrm{i}}$ are temperature dependent; and the numerator and denominator are such that not only their second order, but also their fourth order partial derivatives with respect to ρ equal to zero at a temperature independent density $\rho_0$ , almost equal to the critical one. Thus, outside the critical region the parsimonious empirical P-ρ-T surfaces of real gases should exhibit the global properties of the Riemann-Hugoniot catastrophe and the characteristics of the swallowtail catastrophe, whereby the zero values of the four parameters $a_2, a_4, b_2, and\ b_4$ in (45) are preassigned to be equal zero through the catastrophe theory!

**(ii) Order of the critical point.** The critical point of a catastrophe is a local property with widespread effects. When constructing an analytic ansatz for an empirical equation of state, the assumption about the analytic nature of the critical point is essential, cf. [12]. One can get empirical information about it by analyzing the behavior of the P-ρ isotherms in a wide neighborhood of the critical point of real gases. Thus, twenty years ago there was an extensive research accomplished for diploma thesis, see appendix in [29]. It provides:

a) an example how to get information about zero of multiplicity 4 from an empirical function by using the Weierstrass preparation theorem.

b) the information about critical points of some ninety empirical equations of state, which suggests that the critical point of a real gas is of the fourth order.

**(iii) Empirical properties of P-ρ-T surfaces.** In 1981, Ribarič and Žekš [22] presented examples of real and model fluids, and experimental evidence in support of these two hypotheses:

**i)** The P-ρ-T surface of any real fluid exhibits at least two critical points, one of them possibly in the subcooled or solid state region.
**ii)** Any substance may also exhibit an anisotropic phase of an immediate order between the liquid and solid phases.

## 4. Empirical Propagators

It took some sixty years for theoretical and experimental contributions to the formulation of the Standard Model, dealing also with infinities due to the too slow decrease of Feynman propagators at large momenta. Yet there are no experimental data about the momentum dependence of the Feynman propagators, though they are essential to QFTs.

We would get such data through Feynman propagators modified with empirical convergence factors. One can obtain such empirical propagators by replacing the Feynman propagators with appropriate ansatze and determine *their free parameters* within the Standard Model by data fitting without renormalization. The Standard Model becomes UV finite through empirical propagators. They are empirical substitutes for realistically regularized Feynman propagators conjectured by Pauli some seventy years ago.

Since one obtains Feynman kind of propagators from the Feynman-Stueckelberg solutions to the Euler-Lagrange equations of the QFT free-field Lagrangians, any information about propagators is also information about free Lagrangian of the Standard Model. By enabling novel information from the present experimental data, the convergence factors of empirical propagators would provide a welcome hint about the underlying physics of the Standard Model.

Field-theoretic infinities first encountered in Lorentz's computation of electron have persisted in classical electrodynamics for seventy and in quantum electrodynamics for some thirty-five years. We consider them in our paper "Empirical regularization of field theory infinities: Electrostatic field energy of electron" [25]. Using empirical regularization of Feynman propagators, we proposed to obtain:
**i)** information about the momentum dependence of Feynman propagators,
**ii)** some universal, standard regulating factors of Feynman propagators to get a self-consistent model of quantum scattering of fundamental particles.

We provide therein the following I+1 parameters ansatz for an empirical regulating factor of Feynman propagators:

$$F(z\,;a,b_i,n,I) = \left(\sqrt{\mathrm{m}^2+a}+\sqrt{\mathrm{a}}\right)^{\mathrm{n}+2\mathrm{I}}\left(\sqrt{\mathrm{z}+\mathrm{a}}+\sqrt{\mathrm{a}}\right)^{-n-2\mathrm{I}} a^{-1}\textstyle\sum_{i=0}^{I} b_i(z-\mathrm{m}^2)^i \quad (46)$$

where $a > m^2$, $\mathrm{n} = 1,2,\ldots$, $b_i$ are real parameters, and $b_0 = a$,

We formulated this ansatz to have the following properties:

**[1st]** $F(z)$ is an analytic function of the complex variable z with a branch cut singularity along the segment $\mathrm{z} \le -\mathrm{a}$ of the negative real axis.

It has no insulated singularity because $\mathrm{Re}\sqrt{z} > 0$.

**[2nd]** $F(z)$ is real for all of $\mathrm{z} > -\mathrm{a}$.

**[3rd]** $F(z) = \mathrm{O}(z^{-n/2})$ as $\mathrm{z} \to \infty$.

**[4th]** $sup_{|z|<z_0}|F(z)-1| \to 0$ if $z_0 < \infty$ *and* $\mathrm{a} \to \infty$.

**[5th]** $F(\mathrm{m}^2) = 1$; Residuum of the ansatz $F(k^2)\widetilde{G_F}(\mathrm{k})$ for empirically regulated Feynman propagator $\widetilde{G_F}(\mathrm{k})$ equals the original one.

Suppose we take the simplest ansatz $F(z\,;a,b_i,n,I=0)$ and calculate the empirical value $a_{emp}$ of parameter $a$ by parsimonious data fitting. The values $F_0 = F\left(z=0\,;a_{emp},b_i,n,I=0\right)$ of the empirically determined regulating factor at $z=0$ and the empirical value $a_{emp}$ satisfy relation

$$a_{emp} = \tfrac{1}{4}\mathrm{m}^2 F_0{}^{\mathrm{n}}/(\,F_0{}^{-\mathrm{n}}-1). \quad (47)$$

Thus

$$a_{emp} \to \infty \;\; when \; F_0 \to 1. \quad (48)$$

This suggests that the empirical values of parameters $a$ *and* $b_i$ of the regulating factor $F(z\,;a,b_i,n,I)$ will be vastly different because one expects that the low momentum values of Feynman propagators seem to be adequate.

## 5. Concluding Remarks

**[1st]** As Feynman [37] pointed out already some sixty years ago: "We have a habit in writing articles published in scientific journals to make the work as finished as possible, to cover all the tracks, to not worry about the blind alleys or to describe how you had the wrong idea first, and so on. So there isn't any place to publish, in a dignified manner, what you actually did in order to get to do the work,..." Thus, some vital information is neglected. Trying to

ameliorate the situation, this article is the eighth version of the one that first appeared in arXiv on 6 Oct 2008. In this version, we take account of editorial suggestions and emphasize the importance of parsimonious data fitting.

[2nd] **Summary.** We considered the history of empirical equations and various constructions, to promote interest in them as an initial model of the physical law. To this end, we provide a variety of historical examples and simulate empirical calculations of Planck's law, and van der Waals equation. We emphasize the importance of parsimonious data-fitting, which is a very heuristic subject unfamiliar to theorists. Thereby we provide **a**) Empirical forms of Planck's law, and **b**) Collation of verified symmetries and catastrophes-like properties of empirical P-ρ-T surface of real gases.

[3rd] One may consider as a stepping–stone to a novel gaseous **thermodynamic law** the above mentioned two empirical findings about a real gas : a) the critical point is of the fourth order and b) the isothermal pressure density dependence is an $A_2$ and $A_4$-rational function of density ρ, cf. (45).

[4th] Empirical relationship seems to be a foreign subject to contemporary theoretical physics, not even a wrong one. Otherwise, one cannot understand the fact that for decades hundreds of QFT computations ware performed with propagators that need regularizations without a single try to calculate an empirical substitute for them, though the Standard Model formulas specify its properties. We feel there is no room for this kind of complacency if we really want to come to physics beyond the Standard Model. Consequently, we are **waiting for a latter-day Planck** to empirically regularize Feynman propagators so as to replace a wide variety of regularization schemes, thereby providing a self-consistent perturbative model of quantum scattering and a stepping-stone to physics beyond the Standard Model, see [25].

[5th] We adhere to the following **Feynman's advice** [37] "The only true physical description is that describing the experimental meaning of the quantities in the equation – or better, the way the equations are to be used in describing experimental observations. This being the case perhaps the best way to proceed is to try to guess equations, and disregard physical models or descriptions. For example, McCullough guessed the correct equations for light propagation in a crystal long before his colleagues using elastic models could make head or tail of the phenomena, or again, Dirac obtained his equation for

the description of the electron by an almost purely mathematical proposition. A simple physical view by which all the contents of this equation can be seen is still lacking. Therefore,
*I think equation guessing might be the best method to proceed to obtain the laws for the part of physics which is presently unknown." "... the problem is not to find the best or most efficient method to proceed to a discovery, but to find any method at all.*"
There may be something to the Latin saying, *" Sic parvis magna*".

**[6th]** The farewell essay "**On equation guessing**" points out issues of equation guessing in modeling natural phenomena. It reminds us that empirical equations created by equation guessing inspired the fundamental theories. Equation guessing is heuristics (trial and error): we know no recipe that you could learn; there are only inspiring examples and suggestions from eminent scientists. In equation guessing, both true and wrong answers are welcome; just not even wrong ones are useless. So start making some, see [40].

**[7th]** *"Cum quad habere cupis, noli cessare petendo: arbor nonprimo, sed saepe cadit,"* is a very relevant saying because creating a quality empirical equation is a long lasting process with many failures.
As they say "*Di scentias laboribus vendunt*"

**[8th]** As an ansatz in computational physics, we frequently use the Padé approximants; kind of versatile models well known for making empirical equations. However, to infer natural laws from experimental data, we need various, sophisticated methods, see Schmidt and Lipson [33]. They touch on:
**i)** algorithms and mathematics for understanding dynamical systems,
**ii)** computational methods for data mining and hypothesizing about the physical and chemical relationships of natural systems,
**iii)** tools that allow scientists to test their ideas and frameworks more rapidly by computers to determine if these ideas produce accurate and parsimonious models, laws, and predictions.

**[9th]** In general, while constructing empirical equations it pays to take account of the specific, various **characteristics of the underlying natural process**: the form of the underlying equations, linearity, symmetries, positivity, and finiteness of data, etc. One should take into account also whether one is modeling a start-up, early time dynamics of the considered

process or a long-term dynamic, such as periodic or asymptotic one, see [35]. Preceding used ansatze may give a hint too; the method of unitary regulators [38] suggests invariant Padé approximants as ansatze for the convergence factors of empirical propagators, cf. [39].

[10th] **Stability.** The stability concept often appears among the keywords of papers on the model making of natural phenomena. When constructing empirical formulas there are two cases of primary importance:

**a)** Stability against the inaccuracy of experimental data, i.e. an empirical formula should be able to store and predict the experimental data with an accuracy that is comparable to that of the stored experimental data.

**b)** Structural stability of the considered phenomenon. Therefore, we propose to use an elementary catastrophe in empirical equation of state.

Stewart [32] presents a broadly based discussion of catastrophe theory, placing emphasis on the developmental feedback between the mathematics and its applications, especially in the physical sciences. According to Stewart: “A structurally stable system preserves its basic form when its *equations* are perturbed: it is robust, not only to small changes in initial data, but to small changes in its own specification. There are philosophical reasons to prefer such systems when modeling nature, especially if the model contains parameters, which are to be estimated from observations; Thom (1972) advocated this stance as a guiding principle, though he was by no means the first to do so. It is wise to interpret it with intelligence, and not as an inflexible rule; but when used this way, it can pay dividends.”

**Acknowledgement**

We express our thanks to S. Ribarič, B. Bajc, and to the AJMP reviewer for many helpful comments.

❖ *Any comment, reference, suggestion, or viewpoint is very welcome and will be considered in the in the next version of this article!*

## 6. References

[1] Geocentric model, En. Wikipedia.

[2] Hooke's law, En. Wikipedia.

[3] Ohm's law, En. Wikipedia.

[4] Balmer series, En. Wikipedia.

[5] Rydberg formula, En. Wikipedia.

[6] A. H. Cook, *The observational foundations of physics,* Cambridge University Press, Cambridge (1994).

[7] Planck's law, En. Wikipedia.

[8] Van der Waals equation, En. Wikipedia.

[9] R. Feynman, *The Character of Physical Law*. Modern Library, ISBN 0-679-60127-9 (1965); THE M.I.T. PRESS, PDF.

[10] M. Ribarič and B. Žekš, P,ρ,T surfaces of real gases and butterfly catastrophe, Chem. Phys. **41** (1979), pp. 221-228.
PDF copy Marijan Ribarič (researchgate.net)/Research

[11] M. Veltman, *Diagrammatica,* Cambridge University Press (1994).

[12] A. V. Tatarenko, The Equation of State from a Butterfly Catastrophe, http://dx.doi.org/10.1063/1.3562541 (2011),

[13] M. Ribarič and B. Žekš, Catastrophe Theory and Liquid-Gas Phase Transitions, Fizika, Vol.8, Suppl. (1976), p. 254.
PDF copy Marijan Ribarič (researchgate.net)/Research

[14] M. Ribarič and B. Žekš, Global behavior experimentally determined P,ρ,T surfaces of real gases and cups and butterfly catastrophes, Fizika. - ISSN 0015-3206. **9** (1977), pp. 205-211.
PDF copy Marijan Ribarič (researchgate.net)/Research

[15] M. Ribarič and B. Žekš, Antisymmetry of the experimentally determined chemical potential of real gases and Thom's theory of catastrophes, Fizika. **10** (1978), pp. 203-215.
PDF copy Marijan Ribarič (researchgate.net)/Research

[16] M. Ribarič and B. Žekš, P,ρ,T surfaces of real gases and the Riemann-Hugoniot catastrophe, Chem. Phys. Vol. 34 (1978), pp. 225-230.

[17] M. Ribarič, D. Stojanovski and B. Žekš Investigation of experimentally determined functional dependences by osculating curves, Fizika. **11** (1979), pp. 17-29. PDF copy Marijan Ribarič (researchgate.net)/Research

[18] M. Ribarič and B. Žekš, Antisymmetry of the logarithm of the compressibility factor z(*) of real gases, Cryogenics. - ISSN 0011-2275. Vol. 20 (1980), pp. 451-454.PDF copy Marijan Ribarič (researchgate.net)/Research

[19] M. Ribarič and B. Žekš, High density behaviour of extrapolated P, ρ, T, surfaces of real fluids and butterfly catastrophe, Fizika. **12** (1980), pp. 41-54. PDF copy Marijan Ribarič (researchgate.net)/Research

[20] M. Ribarič and B. Žekš, Characteristics of the P,ρ,T surfaces of real fluids in the region of solid state densities, Fizika. Vol. 13 (1981), pp. 1-14. PDF copy Marijan Ribarič (researchgate.net)/Research

[21] M. Ribarič and B. Žekš, Local antisymmetries of isotermal behaviour of the Lenart-Jones and real fluids, Fizika. - ISSN 0015-3206. Vol. 13 (1981), pp. 221-240 PDF copy Marijan Ribarič (researchgate.net)/Research

[22] M. Ribarič and B. Žekš. Possible universal existence of a mesophase, Fizika. **14** (1982), pp. 29-33. PDF copy Marijan Ribarič (researchgate.net)/Research

[23] M. Ribarič and B. Žekš, Phase transitions of real fluids and Thom's theory of catastrophes : presented at 6th Yugoslav Symposium of the Physics of Condensed Matter, Kruševac 1978, Fizika. - ISSN 0015-3206. Vol. 10, Suppl. 2 (1978), pp. 432-435. PDF copy Marijan Ribarič (researchgate.net)/Research

[24] M. Ribarič and B. Žekš ;On whether P,ρ,T surfaces of real fluids exhibit an additional critical point in the vicinity of the triple point, 7th Yugoslav Symposium of the Physics of Condensed Matter, Ohrid 1980, Fizika. **12**, Supp. 1 (1980), pp. 265-266.PDF copy Marijan Ribarič (researchgate.net)/Research

[25] M. Ribarič and L. Šušteršič, Empirical regularization of field theory infinities: Electrostatic field energy of electron, Second Version Research Gate, 2019. DOI: 10.13140/RG.2.2.21457.15209 PDF copy Marijan Ribarič (researchgate.net)/Research

[26] M. Ribarič, I. Bizjak, M. Bonač, M. Lakner, B. Mitrovič, Z. Radalj, D. Stojanovski, and B. Žekš, *Computational Methods for Parsimonious Data Fitting*, Comstat Lectures Vol.2, Physica Verlag, Wuerzburg-Vienna 1984. 154pp. ISBN 3-7051-0004-1, PDF copy Marijan Ribarič (researchgate.net)/Research

[27] J. Thijssen, *Computational Physics*. Cambridge University Press. (2007).

[28] M. Ribarič and L. Šušteršič, Heuristic key to Physics beyond the Standard Model: Modification of the QFTs so as to make their diagrams convergent, Sects. 5.3.1 and 5,3.2. arXiv:1503.06325 [physics.gen-ph], 2017.

[29] M. Ribarič and L. Šušteršič, Empirical formulas for prediction of experimental data & Appendix, Research Gate, 2017:
PDF copy Marijan Ribarič (researchgate.net)/Research

[30] S. Schulman and M. Revzen, Phase transitions as catastrophes, Collective Phenomena **1,** pp.43-49 (1972).

[31] R. Thom, Structural stability and morphogenesis, W. A. Benjamin, Inc, Reading, Mass. 1975.

[32] I. Stewart, Catastrophe theory in physics, Rep. Prog. Phys., Vol. **45,** 1982.
https://www.researchgate.net/publication/231098402

[33] M. Schmidt and H. Lipson, Distilling Free-Form Natural Laws from Experimental Data, *Science* 03 Apr 2009:Vol. 324, Issue 5923, pp. 81-85
DOI: 10.1126/science.1165893.

[34] H. Kragh, Max Planck: The reluctant revolutionary, Physics World,2000.

[35] M. Ribarič and L. Šušteršič, Simplified models: a novel approach to dynamics of point-like objects, **arXiv:1305.6041v2**, 2016.

[36] C. Baumgarten, The Final Theory of Physics - a Tautology?
arXiv:1702.00301v2 [physics.gen-ph], 2017.

[37] https://www.nobelprize.org/nobel_prizes/physics/laureates/1965/feynman-lecture.html.

[38] G. 't Hooft and M. Veltman, Diagrammar, CERN Yellow Report 1973.

[39] M. Ribarič and L. Šušteršič, Invariances of regularized QED diagrams, 2000arXiv:hep-th/0010209v1

[40] M. Ribarič, *On Equation Guessing*, Research Gate, 2020.
https://www.researchgate.net/publication/340249381

[41] History of calculus - Wikipedia.

[42] https://en.wikipedia.org/wiki/Stefan–Boltzmann_law.